# Voice vs. Data:
# Estimates of Media Usage and Network Traffic

A. Michael Noll

Annenberg School for Communication
University of Southern California
Los Angeles, CA  90089-0281
(908)  647-3294

and

Columbia Institute for Tele-Information
Columbia University
New York, NY  10027-6902
(212)  854-4222

September 2, 2001

ABSTRACT
   This paper reports on the results of surveys of the media usage by two small groups of students, one group in New York City and the other in Los Angeles. The voice and data traffic implications of the results are estimated. When converted to bits, the telephone traffic was much greater than the data traffic. An attempt is made to reconcile the findings with network usage data that implies that voice exceeds data.

INTRODUCTION
   The popular conception is that data traffic nearly, if not already, exceeds voice traffic on backbone networks. However, the results of research reported in this paper imply that user-generated voice traffic exceeds data traffic. This finding is the result of asking real users to estimate their usage of a wide variety of media and then converting media usage to the common measure of bits. Media usage was surveyed for students in New York City and in Los Angeles. Other than significant differences in radio listening, e-mails, and downloads, the media usage was quite similar. Telephone usage (wired and wireless) was nearly an hour per day. When converted to bits, the telephone traffic was much greater than the data traffic over the Internet.

MEDIA USAGE SURVEYS
   Two media usage surveys were conducted: one of MBA students in New York City and the other of undergraduate communications students in Los Angeles. A short, one-page, self-administered questionnaire was distributed to the students. The students were asked to "think about a usual average day within the past 7 days" and then to estimate their "total use for that day for the following media." The media categories were the telephone (excluding their cell phone); the cell phone; the Internet; television; radio; music; and newspaper, magazines, and books. Some of the categories were subdivided. The complete wording of the questionnaire is shown in Table 1.

A total of 29 graduate students in New York City and of 21 undergraduate students in Los Angeles responded to the survey. The average responses for both groups of students are tabulated in Table 2.

SURVEY RESULTS

The graduate business students spent an average of 39.0 minutes speaking on the telephone and 19.1 minutes on the cell phone, for a total voice telecommunications of 58.1 minutes. The undergraduates spent 35.6 minutes on the telephone and 15.5 minutes on the cell phone, for a total voice telecommunications of 51.1 minutes.

The graduate business students sent 16.1 e-mails and received 50.5 e-mails—a ratio of about 3 to 1. The undergraduate students sent 3.6 e-mails and received 10.5 e-mails—again a ratio of about 3 to 1. The graduate business students visited 11.6 Web sites, and the undergraduates visited 7.3 Web sites. The graduate business students had an average of 0.5 software downloads and 0.5 A/V downloads. The undergraduate students had an average of 0.9 software downloads and 4.2 A/V downloads.

The graduate business students watch television 1.5 hours, and the undergraduates watch television 2.0 hours. The graduate business students watch videos for 0.3 hours, and the undergraduates watch videos for 1.2 hours. Radio was listened to for 24.3 minutes by the graduate business students and for 72.6 minutes by the undergraduates. Music is listened to for 74.7 minutes by the graduate business students and for 58.2 minutes by the undergraduates. The total amount of time spent watching and listening to entertainment media is about 3.5 hours for the graduate business students and about 5.4 hours for the undergraduates.

Newspapers, magazines, and books are read by the graduate business students for 36.9, 24.9, and 86.7 minutes respectively. Newspapers, magazines, and books are read by the undergraduate students for 21.2, 16.7, and 81.2 minutes respectively. Total print usage is 148.5 minutes for the graduate students and 119.1 minutes for the undergraduates.

SURVEY DISCUSSION

Most differences between the two groups of respondents were minor in terms of their media usage. The business students received five times as many e-mails as the undergraduates, probably because many of the business students were already working and thus received many e-mails at work.

The undergraduates had nearly five times as many A/V downloads as the business students. The undergraduates were asked as a group what they were downloading. The response was audio Napster-like files. The undergraduates watch videos four times as much as the graduate students. When asked as a group, nearly all the undergraduates had DVD players and clearly were watching them and their VCRs.

The undergraduates listen to radio for three times as much as the graduate students. This is not surprising since the undergraduates live in Los Angeles, which is one of the top radio markets in the United States. People in Los Angeles routinely spend much time commuting and driving in their automobiles, where listening to the radio is a major driving diversion. Radio does not work in the subways of New York City. The undergraduates spent over five hours with entertainment media and the business students spent somewhat less, about 3.5 hours. But both groups clearly like to be entertained by the media during an average day.

Paper-based print media are used somewhat more by the graduate students than by the undergraduates, but both groups spend about two hours with print media. The business students spend more time with magazines and newspapers than the undergraduates. This could be because the business students work and thus must stay current by reading timely material.

VOICE *vs.* DATA

Both the business students and the undergraduate students spoke daily on the telephone for a total of about 60 minutes. Voice is digitized for long-distance at a rate of 64 Kbps in each direction, whether the person is speaking or not. Thus, voice generates 128 Kbps of backbone network traffic. The 60 minutes of telephone usage therefore creates 128 Kbps X 60 sec/min X 60 mins = 460,800 Kbits, or about 460 Mbits.

The e-mail traffic varied greatly. The largest average was for the business students at about 67 e-mails. Assume an average e-mail consists of 2,000 words with an average word consisting of 6 characters at 8 bits per character. This gives a total traffic for e-mail content of 6.4 Mbits. A reasonable average of 10 web sites were visited daily. Assume visiting each web site involves 200 Kbytes. This creates a total web traffic of 16 Mbits. Assume an average daily download of 2 Mbytes. Converting this to bits gives a total of 16 Mbits. The total for all the Internet-related traffic is about 38 Mbits. Even if an overhead for data were encountered equal to twice the traffic, the total traffic would then be only 76 Mbits, still much less than the voice traffic of 460 Mbits.

WEB TRAFFIC

The University of Southern California (USC) computer center provides detailed reports of the traffic on its connection to the Internet. For a typical day (April 30, 2001), the report showed an average traffic of about 17.8 Mbps ingoing and about 10.8 Mbps outgoing for just one of the connections to the Internet. Peak traffic of about 50 Mbps occurred for short bursts around 2 AM. No hourly variation was reported in the traffic, which is fairly uniform, other then the aforementioned peaks. Such a uniform daily traffic pattern is puzzling, since most users are asleep during the early morning hours.

According to the manager of the USC computer center, USC has about 400 two-way voice trunks connecting it to the public switched network. These trunks are capable of carrying 25.6 Mbps in each direction for a total two-way peak traffic of about 50 Mbps. According to the same manager, the total daily average traffic (incoming and outgoing) between USC and the Internet is 120 Mbps. Thus, Internet data traffic is more than twice the voice traffic for USC. This seems puzzling based upon user estimates of media usage which support the opposite finding, namely that voice exceeds data traffic.

The On-line Journalism Review (OJR) maintained by the Annenberg School operates its own server. Detailed reports for this site indicate that a substantial number of visitors are web crawlers and search engines. This finding was also confirmed for the web site maintained by the Columbia Institute for Tele-Information. One crawler, Alexa, reports "gathering in excess of 118 gigabytes of information per day."[1] In July, 2000, the Alexa archive contained 31.6 gigabytes and had doubled since December, 1999.[2] Could it be that a substantial proportion of the traffic on the Internet is generated—not by real users—but by crawlers and search engines?

DISCUSSION

My estimates concluded that voice traffic greatly exceeded data traffic carried over backbone networks, now and for the foreseeable future [Noll, 1991 and Noll, 1999]. This conclusion is controversial and some scholars and industry people disagree with the finding [*Comm. ACM*, 1999]. Although this could just be hype and promotion of corporate positions, the number and variety of people stating that data exceeds voice traffic is cause to give more thought to this question.

The results of the surveys reported in this paper indicate that voice traffic greatly exceeds data traffic. Yet if the amount of data traffic being carried over the Internet indeed exceeds the amount of voice traffic being carried over voice networks, then what is the source of the tremendous amount of data traffic since it apparently is not coming from real human users? One possible source is Web crawlers and search engines. Clearly more study and research is needed to answer this apparent conundrum. Like most research, the answers to one question only raise many more. In the end, the major conclusion is that more data and analysis of the traffic carried over the Web is needed.

---

[1] http://www.alexa.com/company/technology.html

[2] op. cit.

**MEDIA SURVEY**

Thank you for taking the time to complete this short questionnaire. Your individual identity has not been recorded; your responses are private.

The purpose of this survey is to understand your use of communication media.

Please think about a **usual average day** within the past 7 days. Estimate your total use for that **day** for the following media:

Telephone (excluding your cell phone) –
    Total time speaking on the telephone: \_\_\_\_\_\_ mins.
    Number of telephone calls made: \_\_\_\_\_\_.
    Number of telephone calls received: \_\_\_\_\_\_.

Cell Phone –
    Total time speaking on your cell phone: \_\_\_\_\_\_ mins.
    Number of calls made on your cell phone: \_\_\_\_\_\_.
    Number of calls received on your cell phone: \_\_\_\_\_\_.

Internet –
    Number of e-mails sent: \_\_\_\_\_\_.
    Number of e-mails received: \_\_\_\_\_\_.
    Number of Web sites visited: \_\_\_\_\_\_.
    Number of computer software downloads: \_\_\_\_\_\_.
    Number of audio or video downloads: : \_\_\_\_\_\_.

Television –
    Hours watching television (cable, VHF/UHF, satellite) : \_\_\_\_\_\_ hrs.
    Hours watching video tapes or DVDs : \_\_\_\_\_\_ hrs.

Radio –
    Time listening to the radio (home, work, or while traveling): \_\_\_\_\_\_ mins.

Music –
    Time listening to audio CDs or tapes (home, work, or while traveling): \_\_\_\_\_\_ mins.

Newspaper, Magazines, & Books –
    Time reading newspapers: \_\_\_\_\_\_ mins.
    Time reading magazines: \_\_\_\_\_\_ mins.
    Time reading all kinds of books: \_\_\_\_\_\_ mins.

Table I.

MEDIA USAGE

| MEDIUM | MBA (NYC) N = 29 | | UNDERGRADS (LA) N = 21 | |
|---|---|---|---|---|
| | average | $\sigma_N$ | average | $\sigma_N$ |
| TELEPHONE: | | | | |
| time | 39.0 mins | 33.3 | 35.6 mins | 34.0 |
| calls made | 5.2 | 5.3 | 3.2 | 1.7 |
| calls received | 4.9 | 4.5 | 3.7 | 2.2 |
| CELL PHONE: | | | | |
| time | 19.1 mins | 27.8 | 15.5 mins | 14.4 |
| calls made | 4.2 | 5.3 | 4.1 | 4.3 |
| calls received | 3.1 | 4.1 | 2.5 | 2.5 |
| [voice total] | [58.1 mins] | | [51.1 mins] | |
| INTERNET: | | | | |
| e-mails sent | 16.1 | 13.0 | 3.6 | 2.5 |
| e-mails received | 50.5 | 68.2 | 10.5 | 7.2 |
| [ratio] | [3 to 1] | | [3 to 1] | |
| Web sites | 11.6 | 12.7 | 7.3 | 5.4 |
| software downloads | 0.5 | 0.6 | 0.9 | 1.2 |
| A/V downloads | 0.5 | 0.7 | 4.2 | 4.9 |
| TELEVISION: | | | | |
| broadcast | 1.5 hrs | 1.7 | 2.0 hrs | 1.1 |
| videos | 0.3 hrs | 0.6 | 1.2 hrs | 1.5 |
| RADIO: | | | | |
| time | 24.3 mins | 34.8 | 72.6 mins | 42.0 |
| MUSIC: | | | | |
| time | 74.7 mins | 96.0 | 58.2 mins | 49.5 |
| PRINT: | | | | |
| newspapers | 36.9 mins | 52.9 | 21.2 mins | 21.1 |
| magazines | 24.9 mins | 28.6 | 16.7 mins | 15.0 |
| books | 86.7 mins | 100.4 | 81.2 mins | 61.0 |

Table II.
CELL PHONE USAGE
(actual users)

| MEDIUM | MBA (NYC) N = 16 (55%) | | UNDERGRADS (LA) N = 17 (81%) | |
|---|---|---|---|---|
| | average | $\sigma_N$ | average | $\sigma_N$ |
| CELL PHONE: | | | | |
| time | 34.7 mins | 29.4 | 19.1 mins | 13.7 |
| calls made | 7.7 | 4.9 | 5.1 | 4.2 |
| calls received | 5.6 | 4.2 | 3.1 | 2.4 |

Table III.
DAILY TRAFFIC

| SERVICE | DAILY TOTAL TRAFFIC |
|---|---|
| VOICE | 460 Mbits |
| e-mail | 6 Mbits |
| web sites | 16 Mbits |
| downloads | 16 Mbits |
| TOTAL DATA | 38 Mbits |